\documentclass[final,5p,times,twocolumn,numbers]{elsarticle}
\biboptions{sort&compress}

\usepackage{amsmath}
\usepackage{amssymb}
\usepackage{xcolor}
\usepackage{multirow}

\newcommand{\ket}[1]{\vert #1 \rangle}

\newcommand{\Tr}{\mathrm{Tr}}

\journal{Physics Letters A}

\begin{document}

\begin{frontmatter}

\title{A variational coherent quantum annealing protocol}

\author[first]{Nancy Barraza}
\affiliation[first]{organization={Universidad Diego Portales, Facultad de Ingeniería y Ciencias, Instituto de Ciencias Básicas, CEDENNA},%Department and Organization
            addressline={Avenida Ejercito Libertador 441}, 
            city={Santiago},
            postcode={8370191}, 
            state={},
            country={Chile}}

\author[second]{Gabriel Alvarado Barrios}
\affiliation[second]{organization={Kipu Quantum GmbH},%Department and Organization
            addressline={Greifswalderstrasse 212}, 
            city={Berlin},
            postcode={10405}, 
            state={},
            country={Germany}}

\author[second]{Enrique Solano}

\author[third]{Francisco Albarrán-Arriagada}
\affiliation[third]{organization={Universidad de Santiago de Chile (USACH), Departamento de Física, CEDENNA},
            addressline={Avenida Víctor Jara 3493}, 
            city={Santiago},
            postcode={9170124}, 
            state={},
            country={Chile}}

\begin{abstract}
We introduce variational coherent quantum annealing (VCQA), a hybrid protocol that optimizes finite-time annealing schedules within a low-dimensional parameter space. We consider three schedule functions that interpolate between the initial, problem (final), and auxiliary Hamiltonians, each represented by a piecewise cubic Hermite interpolant. The auxiliary Hamiltonian is a problem-informed, longitudinal, and local field that vanishes at both ends of the evolution, providing a mechanism to enlarge the minimum energy gap. We benchmark our protocol on QUBO and (HUBO) instances, as well as, a non-stoquastic model, using systems of up to ten qubits. With only six variational parameters for the entire protocol, we find that the characteristic time of the adiabatic-like evolution is reduced by at least a factor of 2.4 in every case. These finite-size, closed-system results establish VCQA as a compact control ansatz and motivate tests under hardware noise and control constraints.
\end{abstract}


\end{frontmatter}

%\tableofcontents

%% \linenumbers

%% main text

\section{Introduction}
Adiabatic quantum computing (AQC) \cite{Albash2018RevModPhys} is a universal computational paradigm, equivalent to gate-based quantum computing in terms of theoretical power~\cite{Aharanov2008SIAMRev}, and it has yielded strong theoretical results~\cite{Childs2001PhysRevA, Sarandy2005PhysRevLett, Hastings2021Quantum}. Its implementation, however, remains challenging due to its analog nature, limited connectivity, and the low coherence times of current devices. Quantum annealing (QA) is a restricted form of AQC devoted to optimization problems, following an adiabatic path from a known initial ground state to the ground state of a problem Hamiltonian that encodes the solution to the optimization task.

Quantum annealers, the experimental devices implementing QA, are mainly designed to solve spin-glass-type Hamiltonians~\cite{Katzgraber2015PhysRevX}, since a broad class of combinatorial optimization problems can be written in the quadratic unconstrained binary optimization (QUBO) formalism~\cite{Lucas2014FrontPhys} and mapped onto a spin glass via a binary-to-spin transformation. This has enabled applications in quantum finance~\cite{Orus2019RevPhys, Orus2019PhysRevA, Ding2023Entropy}, chemistry~\cite{Xia2018JPhysChemB, Du2010PhysRevLett}, machine learning~\cite{Hu2019QuantumEng, Date2021SciRep, Willsch2020CompPhysCommun, Nath2021SNCompSci}, factorization~\cite{Peng2008PhysRevLett}, and logistics~\cite{Weinberg2023SciRep}. Superconducting circuits remain a leading platform due to their scalability and flexibility~\cite{King2022NatPhys, King2023Nature}, alongside recent developments in neutral-atom arrays~\cite{Finvzgar2024PhysRevRes}.

Despite this versatility and its quantum features~\cite{Boixo2016NatCommun, Lanting2014PhysRevX, Albash2015PhysRevA}, whether QA provides a genuine quantum advantage remains an open question~\cite{Amin2015PhysRevA, Koshikawa2021JPhysSocJpn}, illustrated by a recent claim of computational advantage in quantum spin-glass simulation~\cite{King2025Science} that was subsequently challenged by classical tensor-network methods~\cite{Tindall2026Science}. Beyond this debate, QA has shown promise for oracular problems and machine-learning applications~\cite{Somma2012PhysRevLett, Simen2025arXiv}. A major limitation is that fully adiabatic evolution requires runtimes that typically exceed the coherence time of current hardware, so the final state may differ substantially from the problem ground state; since the adiabatic time is tied to the first energy gap, the evolution is especially slow in large or frustrated systems, where the spectrum tends to squeeze. Shortcuts to adiabaticity (STA)~\cite{GueryOdelin2019RevModPhys} address this limitation through techniques such as fast-forward~\cite{Masuda2022PTRSA}, counter-diabatic driving~\cite{DelCampo2013PhysRevLett, Nakahara2022PTRSA}, Lyapunov control~\cite{Ran2017PhysRevA, Chandarana2024arXiv}, variational approaches~\cite{Passarelli2022PhysRevA, Sun2022PTRSA, Barraza2022QuantumSciTechnol}, and digitized adiabatic quantum computing~\cite{Hegade2021PhysRevAppl}, all of which have shown promise for maintaining high-fidelity evolution within the coherence budget of current technology.

In this work, we introduce variational coherent quantum annealing (VCQA), a hybrid classical-quantum protocol designed to operate within the coherence times of current quantum annealers. The schedule functions governing the interpolation between the initial and problem Hamiltonians are expressed as piecewise-continuous functions, built using a polynomial cubic Hermite interpolator (PCHIP) through variational points that ensure continuity of the function and its derivative and monotonic behavior between points; their parameters are optimized via classical routines. The number of variational parameters can remain constant even as the problem complexity grows, keeping the search efficient and well suited to short-time coherent evolutions. VCQA further introduces an auxiliary Hamiltonian that vanishes at the initial and final times and modifies the energy landscape during the evolution, helping to suppress unwanted level crossings and near-degeneracy regions, improving the final ground-state fidelity even at short evolution times~\cite{AlbarranArriagada2026arXiv}. This auxiliary term is fundamentally different from counterdiabatic driving, since it does not mimic the ground state of any part of the total Hamiltonian.

Other approaches introduce auxiliary terms in adiabatic protocols, including non-stoquastic antiferromagnetic drivers, reverse annealing with local bias fields, and diagonal catalysts switched on and off during the interpolation~\cite{Seki2012PhysrevE,Ohkuwa2018PhysRevA,Albash2021PhysRevA}. VCQA instead employs a simple auxiliary Hamiltonian of local operators tied directly to the problem Hamiltonian, combined with cubic Hermite-parametrized schedules -- a problem-informed, hardware-efficient, low-parameter ansatz well suited to near-term gate-based devices. Genetic optimization of annealing schedules combined with local counterdiabatic operators has also been explored~\cite{Hegde2022PhysRevA}; VCQA follows a different route, replacing evolutionary optimization of high-order polynomials with a compact variational ansatz and a third, auxiliary schedule function, balancing expressivity, stability, and experimental feasibility.

We apply VCQA to QUBO-type problems including linear, cyclic, and star Ising connectivities and the weighted MaxCut; as well as a four-body (HUBO) model and a non-stoquastic Heisenberg chain, whose Hamiltonian has both positive and negative off-diagonal elements in the computational basis and is therefore affected by the Monte Carlo sign problem; although not implementable on current QA hardware, it illustrates the broader reach of VCQA beyond stoquastic models. We show that a small number of variational parameters can significantly reduce the final ground-state error across these benchmarks, supporting short coherent evolutions enhanced by optimized controls and auxiliary terms as a viable path for near-term quantum optimization.

\section{Methodology}
We consider a general adiabatic evolution described by the time-dependent Hamiltonian,
\begin{equation}
H(t)=F_1(t/T) H_i + F_2(t/T) H_f + F_3(t/T) H_{aux},
\label{Eq01}
\end{equation}
with $F_1(0)=F_2(1)=1$, and $F_1(1)=F_2(0)=F_3(0)=F_3(1)=0$. $H_i$ is the initial Hamiltonian whose ground state is easy to prepare. In general, for quantum annealers, $H_i$ takes the form
\begin{equation}
H_i=-\epsilon\sum_{j=1}^N\sigma^x_j,
\label{Eq02}
\end{equation}
whose ground state is $\ket{\psi(0)}=\bigotimes_{j=1}^N\ket{+}_j$, where $\ket{+}_j=(\ket{0}_j+\ket{1}_j)/\sqrt{2}$, that is the flat superposition of all the states in the $z$-basis. $H_f$ is the final Hamiltonian whose ground state codifies the solution of an optimization problem. In general, for quantum annealers $H_f$ corresponds to a spin-glass Hamiltonian given by
\begin{equation}
H_f=\sum_{j=1}^N\omega_j\sigma^z_j+\sum_{j,k;j>k}g_{j,k}\sigma^z_j\sigma^z_k,
\label{Eq03}
\end{equation}
whose ground state is a separable state in the $z$-basis, that is, a binary string.

Finally, $H_{aux}$ is an auxiliary Hamiltonian that serves a crucial role in our VCQA protocol by altering the energy landscape during the evolution process to optimize performance. The specific form of $H_{aux}$ will be discussed later.

To optimize the functions $F_1$, $F_2$, and $F_3$, we use a parametrization approach based on cubic piece-wise interpolation through mobile points, which play the role of variational parameters. As indicated in Eq. (\ref{Eq01}), we need to parametrize three functions. If we use $n_j$ parameters for the function $F_j$, then our cost function will depend on $\sum_{j=1}^{3} n_j$ parameters.

\subsection{Parametrization of the schedule function}

To parametrize each schedule function $F_j$ we introduce $N_j$ variational parameters $\vec{p}_j=(p_{j,1},p_{j,2},\dots,p_{j,N_j})$. Together with the two boundary values $p_{j,0}$ and $p_{j,f}$, fixed by the boundary conditions on $F_j(0)$ and $F_j(1)$ respectively, these parameters define a set of $N_j+2$ points $\mathcal{P}=\{(0,p_{j,0}),(\Delta_j,p_{j,1}),(2\Delta_j,p_{j,2}),\dots, (1,p_{j,f})\}$, where $\Delta_j=1/(N_j+1)$; for simplicity, we choose these points to be equally spaced. The $N_j+2$ points of $\mathcal{P}$ can define $N_j+1$ piecewise functions, which we build using a piecewise cubic Hermite interpolator.

We require this interpolator to satisfy three conditions so that the resulting schedule functions remain experimentally feasible. That is, continuity of the function and of its first derivative (smoothness), so that $F_j$ does not exhibit abrupt changes, making it more reasonable for implementation. Also, we request that the schedule function must not exceed a maximum value or fall below a minimum one, with these boundary values set by the experimental capabilities of the physical device intended to generate the schedule. We enforce this by imposing a monotonic condition between adjacent points of the interpolator, so that any bound on $F_j$ can be addressed simply by bounding the points (parameters) $p_{j,i}$ during the interpolation process.

The piecewise cubic Hermite interpolator satisfies these conditions~\cite{FritschCarlson1980}, and is constructed as follows. Using the set of points $\mathcal{P}$, we define the schedule function $F_j$ as
\begin{eqnarray}
F_j(x)=&&h_{j,k,0}(x)p_{j,k} + h_{j,k,1}(x)p_{j,k+1} + h_{j,k,2}(x)\Delta_jm_{j,k}\nonumber\\
&& + h_{j,k,3}(x)\Delta_jm_{j,k+1},\,
\label{Eq04}
\end{eqnarray}
with $k\Delta_j\le x\le (k+1)\Delta_j$. The functions $h_{j,k,\ell}$ are given by
\begin{eqnarray}
h_{j,k,0}&=&\left[1+2\left(\frac{x-k\Delta_j}{\Delta_j}\right)\right]\left[1-\left(\frac{x-k\Delta_j}{\Delta_j}\right)\right]^2,\nonumber\\
h_{j,k,1}&=&\left(\frac{x-k\Delta_j}{\Delta_j}\right)^2\left[3-2\left(\frac{x-k\Delta_j}{\Delta_j}\right)\right],\nonumber\\
h_{j,k,2}&=&\left(\frac{x-k\Delta_j}{\Delta_j}\right)\left[1-\left(\frac{x-k\Delta_j}{\Delta_j}\right)\right]^2,\nonumber\\
h_{j,k,3}&=&\left(\frac{x-k\Delta_j}{\Delta_j}\right)^2\left[1-\left(\frac{x-k\Delta_j}{\Delta_j}\right)\right],
\label{Eq05}
\end{eqnarray}
and the slopes $m_{k}$ at $x=k\Delta_j$ and $m_{k+1}$ at $x=(k+1)\Delta_j$ are selected according to
\begin{equation}
m_{j,k}=\frac{2(p_{j,k+1}-p_{j,k})(p_{j,k+2}-p_{j,k+1})}{\Delta_j(p_{j,k+2}-p_{j,k})}
\label{Eq06}
\end{equation}
if and only if $(p_{j,k+1}-p_{j,k})(p_{j,k+2}-p_{j,k+1})>0$, otherwise $m_{j,k}=0$. As stated above, this construction guarantees $p_{min}<F_j(x)<p_{max}$, where $p_{min}$ ($p_{max}$) is the minimum (maximum) among all the points $p_{j,k}$, related to the range of values the experimental device can generate.

\subsection{Auxiliary Hamiltonian}
This additional term modifies the energy spectrum of the Hamiltonian throughout the evolution. Because we want a protocol implementable on essentially any annealing platform, local terms are the natural choice: a single-qubit field is available wherever the qubits themselves are controlled. In the spin glass Hamiltonian each spin has distinct frequency $\omega_j$, so we take the auxiliary Hamiltonian to be built from local terms of the form
\begin{equation}
H_{aux}=\sum_j\omega_j\sigma_j^{\alpha},
\label{Eq07}
\end{equation}
\begin{figure}[t]
\centering
	\includegraphics[width=1\linewidth]{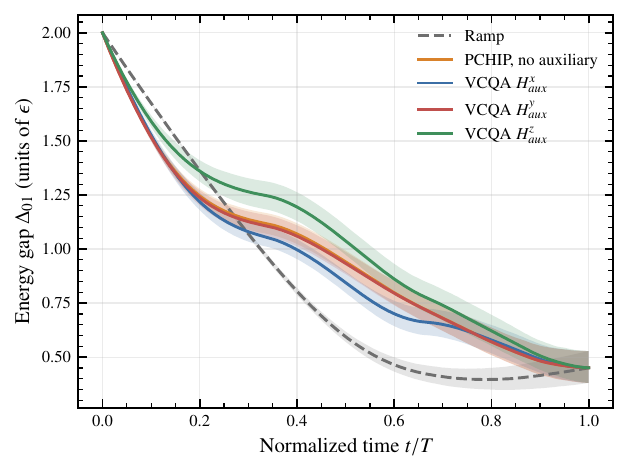}
	\caption{Energy Gap $\Delta_{01}$ as a function of normalized time, for different auxiliary Hamiltonians.}
	\label{Fig01}
\end{figure}
where $\alpha=\{x,y,z\}$. As a first step, we test which choice of $\alpha$ enlarges the energy gap the most. For this purpose we use a small testbed: a fully connected 6-qubit toy problem, based on the Hamiltonian of Eq.~(\ref{Eq03}), with $\omega_j$ and $g_{j,k}$ drawn from $[0,1]$ across 100 random instances.

Fig.~\ref{Fig01} shows the average energy gap $\Delta_{01}$ between the ground and first excited state for each strategy. First, the standard ramp schedule, $F_1=1-t/T$, $F_2=t/T$, and $F_3=0$ (dashed line). Second, the optimized schedule without an auxiliary Hamiltonian (orange line), using two parameters per schedule function, four in total. Finally, the optimized schedules with an auxiliary Hamiltonian of local operators along the $x$-axis (blue line), $y$-axis (red line), and $z$-axis (green line), in each case considering two parameters per schedule function, six in total. Adding the auxiliary Hamiltonian increases the average minimum gap during the evolution and reduces the incidence of near-degeneracy regions, confirming the intuition that motivated considering local terms in the first place. For very short annealing times, however, non-adiabatic transitions can still arise, which limits the achievable performance even with an enlarged gap.

The ramp strategy typically shows a minimum gap at around 0.8 in normalized time before increasing again, a dip that the other strategies avoid. Optimized schedules for local interactions along the $x$ and $y$ axes, as well as schedules without an auxiliary term, show an initial region where the gap falls below that of the ramp; this reflects the limited number of parameters available to the optimizer in these cases. The $z$-local auxiliary Hamiltonian consistently outperforms the other strategies in enlarging the energy gap, and, like most of the optimized strategies, keeps the gap monotonic throughout the evolution. Among local choices for the auxiliary Hamiltonian, this makes $z$-local the best option.

In the $z$-local strategy, the frequency of each qubit in the auxiliary Hamiltonian matches the local term of the final Hamiltonian, which lets the local and bi-local terms evolve effectively under independent schedule functions. On a programmable quantum annealer, controlling local Hamiltonian terms independently of the bi-local ones appears to be a feasible requirement, and this independence is what drives the marked gain in gap size across the evolution, shortening the total time needed to reach a given precision.

Therefore, using the local part of the final Hamiltonian as our auxiliary Hamiltonian, we can write the time-dependent Hamiltonian as
\begin{eqnarray}
H(t)=&&-F_1\epsilon\sum_{j}\sigma_j^x + (F_2+F_3)\sum_{j}\omega_j\sigma^z_j \nonumber\\
&&+F_2\sum_{j,k;j>k}g_{j,k}\sigma^z_j\sigma^z_k
\label{Eq08}
\end{eqnarray}
where $F_j\equiv F_j(t/T)$.
\begin{figure}[t!]
\centering
	\includegraphics[width=0.9\linewidth]{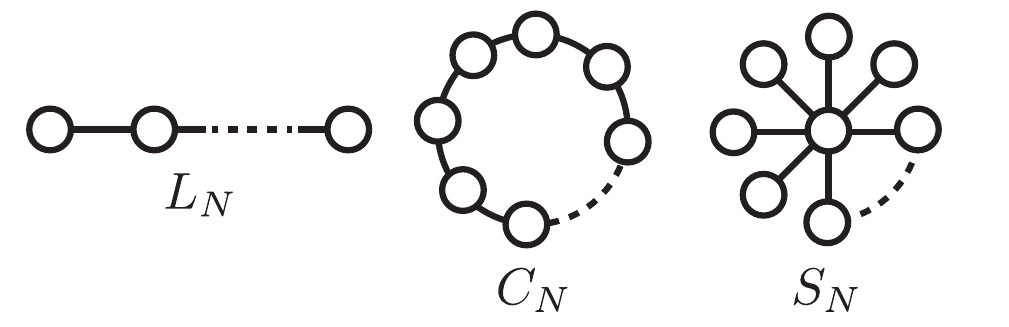}
	\caption{Depiction of the graphs used to illustrate the connectivities of the spin glass Hamiltonians in the considered examples.}
	\label{Fig02}
\end{figure}
As anticipated in Ref.~\cite{AlbarranArriagada2026arXiv} and consistent with the enlarged average gap observed in Fig.~\ref{Fig01}, local terms tend to move the system away from degeneracy points during the quantum evolution. These terms add energy to specific states in the computational basis, biasing the system toward certain configurations in scenarios where degeneracy might otherwise occur, and can be implemented across a variety of quantum computing platforms through local driving fields.

To show the efficacy of our algorithm, we test it on spin glass Hamiltonians with various types of connectivity: linear ($L_N$), cyclic ($C_N$), and star ($S_N$) arrangements, illustrated in Figure~\ref{Fig02}, along with random instances of the weighted maxcut problem. We also report the performance of VCQA in finding the ground state of a non-stoquastic Hamiltonian; although this case cannot be implemented on current quantum annealers, it is interesting because its ground state carries non-trivial quantum correlations. Computational limitations restrict our study to systems of up to 10 qubits. For this evaluation we use the $z$-local auxiliary Hamiltonian described above.

In contrast to other approaches, such as Ref.~\cite{Hegde2022PhysRevA}, where annealing schedules are optimized through genetic algorithms applied to polynomial expansions combined with variationally constructed local driving operators, VCQA adopts a different variational structure: an auxiliary Hamiltonian acting as an effective bias field, together with three schedule functions parametrized by cubic Hermite interpolators. This ansatz avoids oscillatory behavior, guarantees monotonicity, and allows analytical control of the schedules, while requiring only a small number of parameters to reach high performance, properties that favor scalability and eventual hardware implementation.

As we mentioned VCQA auxiliary Hamiltonian contains only local $\sigma_z$ fields that are available in existing annealing architectures, and related mean-field counterdiabatic protocols have been mapped to programmable D-Wave controls~\cite{Hayasaka2023arXiv}. Implementing Eq.~(\ref{Eq08}), however, also requires sufficiently independent and time-resolved control of longitudinal fields and couplers. This work does not model a specific control stack, calibration error, finite temperature, or decoherence, so hardware feasibility should be read as motivation for future experiments rather than as a demonstrated capability.

\subsection{Metrics}
We evaluate performance with two metrics. First, the normalized energy distance $\mathcal{D}$,
\begin{equation}
\mathcal{D}=\frac{\langle\psi(0)|H_f|\psi(0)\rangle-\langle\psi(T)|H_f|\psi(T)\rangle}{\langle\psi(0)|H_f|\psi(0)\rangle - E_{\textrm{ground}}},
    \label{Eq09}
\end{equation}
where $|\psi(t)\rangle$ is the state of the system at time $t$, and $E_{\textrm{ground}}$ is the ground energy of $H_f$. This metric measures how much the expectation value of $H_f$ was reduced, normalized to the maximal reduction: in the worst case $\mathcal{D}=0$ (initial condition), in the best case $\mathcal{D}=1$ (ground state achieved). Second, the ground subspace probability $\mathcal{P}_{\rm GS}$,
\begin{equation}
    \mathcal{P}_{\rm GS}=\sum_{j=1}^{d_{\rm GS}}\langle\psi(T)|\Psi^{\rm GS}_j\rangle\langle \Psi^{\rm GS}_j|\psi(T)\rangle,
    \label{EqPGS}
\end{equation}
where $d_{\rm GS}$ is the degeneracy of the ground state and $|\Psi^{\rm GS}_j\rangle$ is the $j$th basis state spanning the ground subspace. The two metrics capture different failure modes: $\mathcal{D}$ can take large values even when the eigenstates of $H_f$ close to $E_{\rm ground}$ are near-degenerate, a case that does not necessarily yield a large $\mathcal{P}_{\rm GS}$.

\subsection{Computational details}
In every case, we use two parameters per schedule function, so the optimization involves only six parameters in total. These parameters were optimized with a hybrid global--local strategy: for the first instance of each benchmark family we run the Basin--Hopping global optimizer with 150 iterations from a random cold start, while subsequent instances are optimized with only 10 Basin--Hopping iterations, warm-started from the optimal parameters of the previous instance. Sequential Least Squares Programming (SLSQP) serves as the local minimizer in both cases, converging to a tolerance of $10^{-5}$; both routines are provided by the \texttt{scipy.optimize} library~\cite{SciPy}. The variational parameters are bounded to the interval $(0,1)$ throughout the optimization. Time evolution under Eq.~(\ref{Eq08}) is computed with QuTiP's \texttt{sesolve} routine~\cite{Johansson2012CPC,Johansson2013CPC}.

\section{Results}
In this section, we present the numerical results to show the performance of our VCQA algorithm in relation to annealing time, using the optimized schedule functions ($F_1$, $F_2$, and $F_3$). Our analysis is centered on spin glass Hamiltonians, specifically the $L_N$, $C_N$, and $S_N$ configurations, see Figure~\ref{Fig02}. We consider 100 random instances for the local energy $\omega_j$ and coupling terms $g_{j,k}$, as defined in Eq.~(\ref{Eq03}), for systems of up to 10 qubits. We then consider a 10-qubit weighted max-cut problem (100 random instances), a 10-qubit four-spin Hamiltonian, and a 10-qubit non-stoquastic Heisenberg chain, a model known to have degeneracy points~\cite{AlbarranArriagada2026arXiv}. Performance under each protocol is quantified using the normalized energy distance $\mathcal{D}$ and the ground-subspace probability $\mathcal{P}_{\rm GS}$ defined previouslly, and we compare all results against the standard ramp schedule function. The following subsections report these curves for each benchmark; and the last one provides a detailed analysis about characteristic annealing time for direct comparison.

\subsection{$L_N$ connectivity}
\begin{figure}[t]
\centering
	\includegraphics[width=1\linewidth]{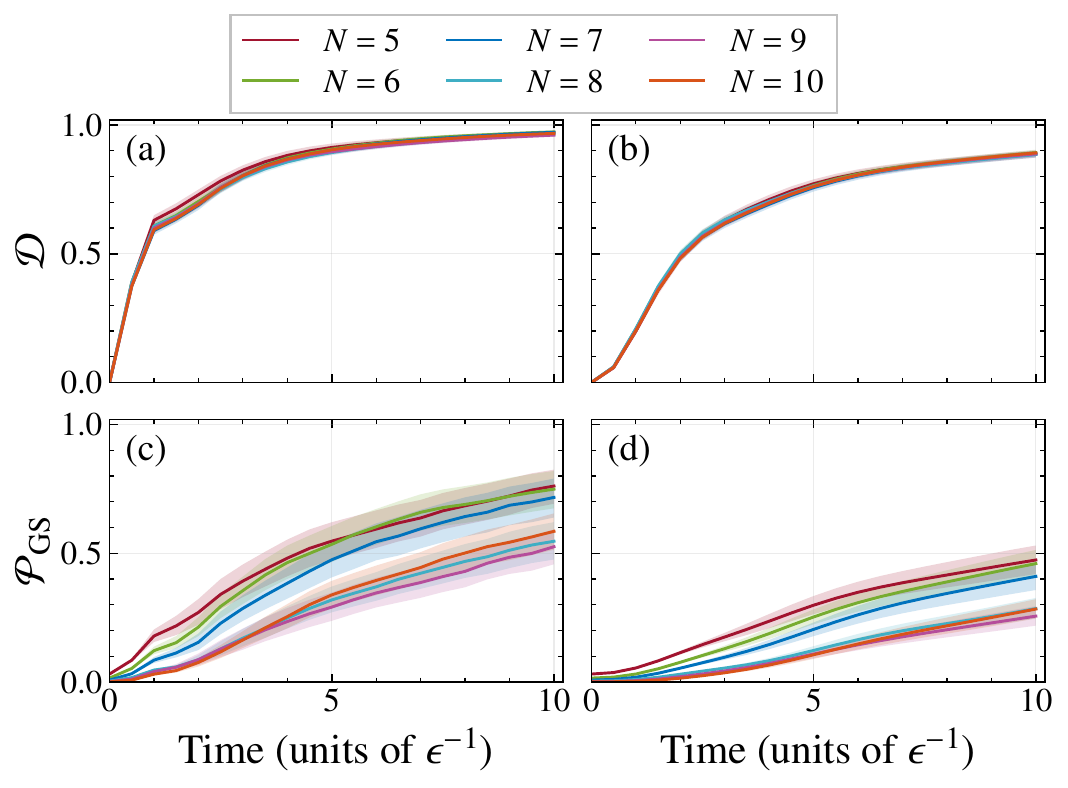}
	\caption{Performance of the VCQA solution as a function of the annealing time for the $L_N$ connectivity, averaged over 100 random instances of $\omega_j,g_j$ for $N=5,\ldots,10$ qubits (shaded bands indicate the spread across instances). Panels (a) and (c) show $\mathcal{D}$ and $\mathcal{P}_{\rm GS}$, respectively, for the VCQA protocol using two parameters per schedule function (six in total). Panels (b) and (d) show $\mathcal{D}$ and $\mathcal{P}_{\rm GS}$, respectively, for the standard ramp schedule function.}
	\label{Fig03}
\end{figure}

As a first example we consider a spin chain with nearest-neighbour interaction, \textit{i.e.} $L_N$ connectivity, described by
\begin{equation}
H_{L_N}=\sum_{j=1}^N\omega_j\sigma^z_j+\sum_{j=1}^{N-1}g_{j}\sigma^z_j\sigma^z_{j+1},
\label{Eq10}
\end{equation}
where $\omega_j$ and $g_j$ are randomly selected within the range $[0,1]$. Figure~\ref{Fig03}~(a) presents the average performance of our VCQA protocol over 100 random instances for $N=5$ to $10$ qubits. The normalized energy distance $\mathcal{D}$ reaches values between approximately 0.95 and 1 within a time of $10\epsilon^{-1}$, with weak dependence on system size. Figure~\ref{Fig03}~(b) shows the same quantity for the ramp schedule function, which saturates at a lower value, around 0.9, over the same time window. In the case of superconducting circuits, $\epsilon$ is on the order of a few gigahertz~\cite{Johnson2011Nature}, so this timescale corresponds to a few nanoseconds.

The advantage of VCQA for probability $\mathcal{P}_{\rm GS}$ is shown in Fig.~\ref{Fig03}~(c) and (d), which plays the role of a generalized fidelity when the ground state may be degenerate. Under VCQA, $\mathcal{P}_{\rm GS}$ reaches values between approximately 0.5 (for $N=10$) and 0.8 (for $N=5$) at $t=10\epsilon^{-1}$, with larger systems attaining both lower values and a wider instance-to-instance spread. Under the ramp schedule function, $\mathcal{P}_{\rm GS}$ remains substantially lower, reaching at most approximately 0.5 for $N=5$ and falling to around 0.25 for $N=10$. 

\subsection{$C_N$ connectivity}
\begin{figure}[t]
\centering
	\includegraphics[width=1\linewidth]{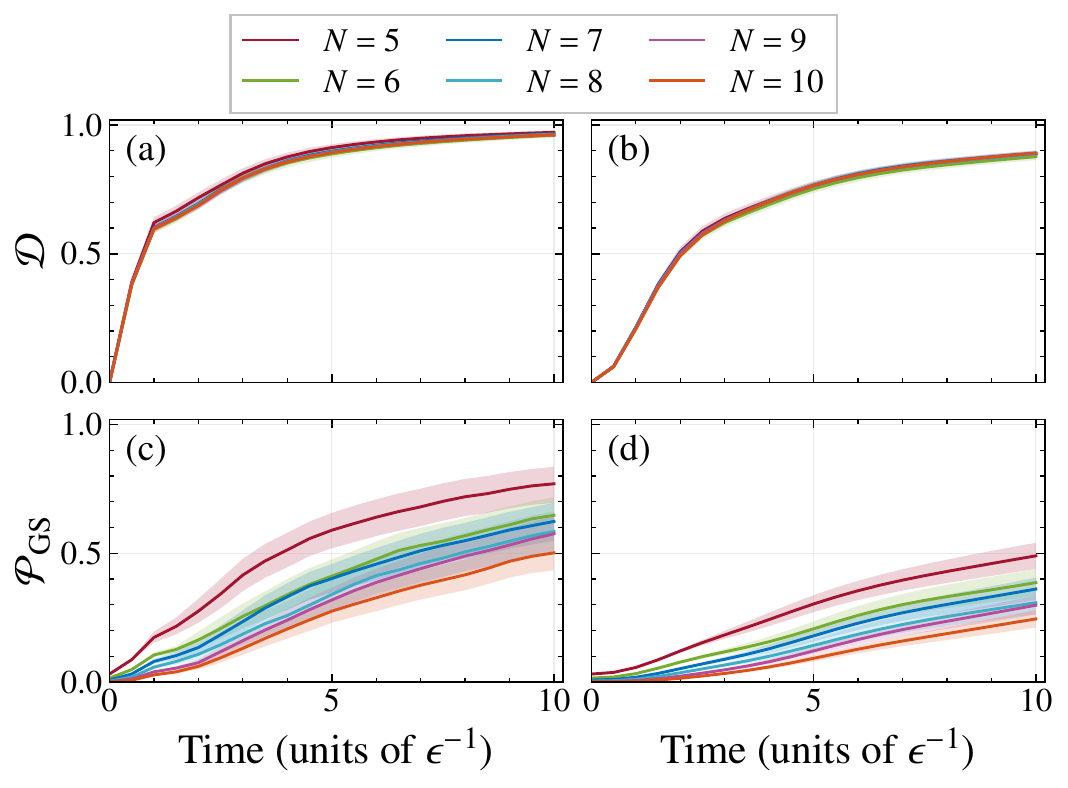}
	\caption{Performance of the VCQA solution as a function of the annealing time for the $C_N$ connectivity, averaged over 100 random instances of $\omega_j,g_j$ for $N=5,\ldots,10$ qubits (shaded bands indicate the spread across instances). Panels (a) and (c) show $\mathcal{D}$ and $\mathcal{P}_{\rm GS}$, respectively, for the VCQA protocol using two parameters per schedule function (six in total). Panels (b) and (d) show $\mathcal{D}$ and $\mathcal{P}_{\rm GS}$, respectively, for the standard ramp schedule function.}
	\label{Fig04}
\end{figure}
Now, we consider a one-dimensional spin chain with periodic boundary conditions, \textit{i.e.} $C_N$ connectivity, described by the Hamiltonian
\begin{equation}
H_{C_N}=\sum_{j=1}^N\omega_j\sigma^z_j+\sum_{j=1}^{N-1}g_{j}\sigma^z_j\sigma^z_{j+1} + g_N\sigma^z_{N}\sigma^z_{1}.
\label{Eq11}
\end{equation}
The Hamiltonians $H_{L_N}$ and $H_{C_N}$ differ only in their last term, specifically the interaction between the first and last site of the chain. This additional term carries significant physical implications, being responsible for the frustration phenomena observed in antiferromagnetic rings with an odd number of sites. As in the previous case, $\omega_j$ and $g_j$ are randomly selected within the range $[0,1]$, and we consider the average behaviour over 100 random instances for $N=5$ to $10$ qubits. Figure~\ref{Fig04} displays the performance of our VCQA algorithm for the $C_N$ configuration. As depicted in Fig.~\ref{Fig04}~(a), $\mathcal{D}$ reaches values close to $0.95$--$1$ within a time frame of $10\,\epsilon^{-1}$, essentially matching the $L_N$ case. The ramp schedule function, shown in Fig.~\ref{Fig04}~(b), again saturates lower, around $0.9$.

As depicted in Fig.~\ref{Fig04}~(c), the ground-subspace probability $\mathcal{P}_{\rm GS}$ ranges from approximately $0.5$ ($N=10$) to $0.8$ ($N=5$) for the VCQA protocol, a spread very similar to that found for $L_N$ connectivity. In contrast, Fig.~\ref{Fig04}~(d) shows that the ramp schedule function only reaches $\mathcal{P}_{\rm GS}$ values around $0.5$ for $N=5$, falling to approximately $0.25$--$0.3$ for $N=10$. This suggests that, at least for these metrics, the additional frustration introduced by the periodic boundary term does not substantially change the relative advantage of VCQA over the ramp strategy observed for the $L_N$ connectivity.

\subsection{$S_N$ connectivity}
\begin{figure}[t]
\centering
	\includegraphics[width=1\linewidth]{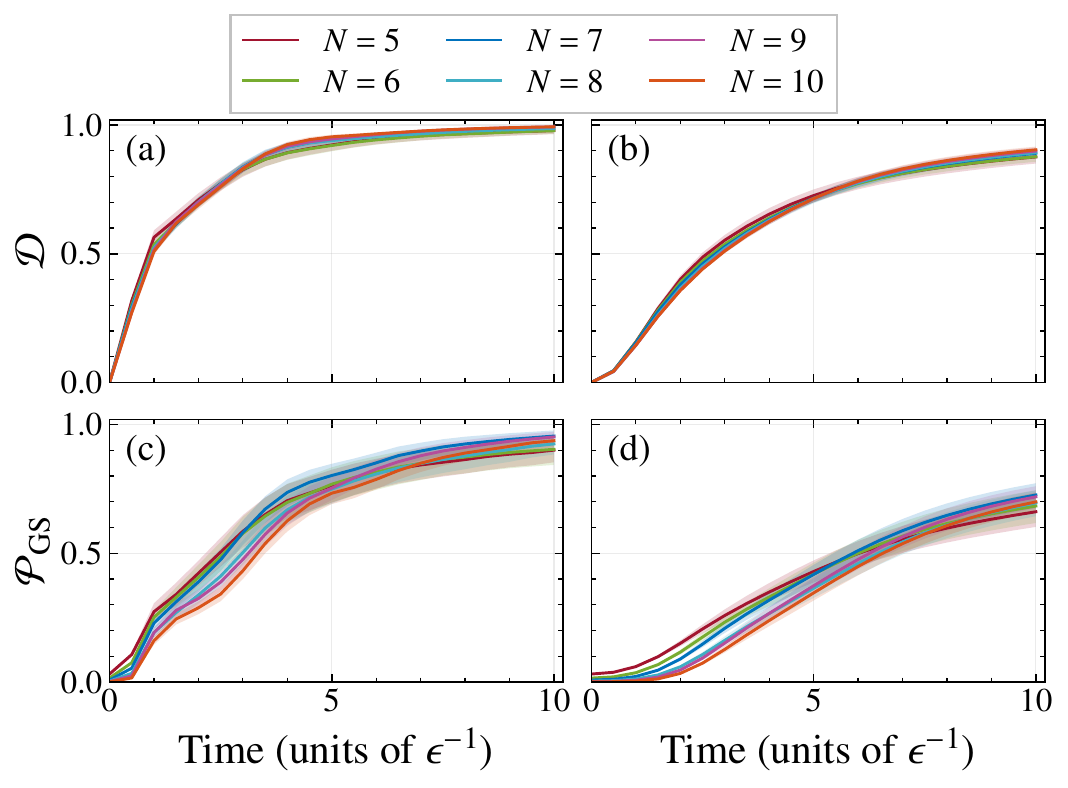}
	\caption{Performance of the VCQA solution as a function of the annealing time for the $S_N$ connectivity, averaged over 100 random instances of $\omega_j,g_j$ for $N=5,\ldots,10$ qubits (shaded bands indicate the spread across instances). Panels (a) and (c) show $\mathcal{D}$ and $\mathcal{P}_{\rm GS}$, respectively, for the VCQA protocol using two parameters per schedule function (six in total). Panels (b) and (d) show $\mathcal{D}$ and $\mathcal{P}_{\rm GS}$, respectively, for the standard ramp schedule function.}
	\label{Fig05}
\end{figure}
Now we consider $S_N$ connectivity, which means that one site is connected to all the others. This situation is described by
\begin{equation}
H_{S_N}=\sum_{j=1}^N\omega_j\sigma^z_j+\sigma^z_1\sum_{j=2}^{N}g_{j}\sigma^z_j.
\label{Eq12}
\end{equation}
This connectivity has the capacity to correlate all the sites through the central one, thereby increasing the multipartite quantum correlations within the system. This characteristic sets this system apart from the other examples, making it a valuable test case. Figure~\ref{Fig05} shows the performance of our proposal for the $S_N$ case involving up to $10$ qubits. Again, the local energies and couplings were selected randomly, where we used 100 instances and plotted the average performance.

As observed in Fig.~\ref{Fig05}~(a), $\mathcal{D}$ under VCQA reaches values close to $0.95$--$1$ within $10\,\epsilon^{-1}$, comparable to the $L_N$ and $C_N$ cases. The ramp schedule function, shown in Fig.~\ref{Fig05}~(b), again saturates around $0.9$.

The $S_N$ connectivity shows the clearest advantage in the ground-subspace probability. As shown in Fig.~\ref{Fig05}~(c), VCQA reaches $\mathcal{P}_{\rm GS}$ values between approximately $0.85$ and $0.95$ for all system sizes considered, with comparatively little spread across $N$. For the ramp schedule function, Fig.~\ref{Fig05}~(d) shows a much wider range, from approximately $0.65$ for $N=10$ up to approximately $0.95$ for $N=5$; in this connectivity the ramp strategy remains competitive with VCQA for small $N$, but its performance degrades markedly as $N$ increases, whereas VCQA remains comparatively stable across the sizes studied.

Across the three connectivities, the normalized energy distance $\mathcal{D}$ shows only weak dependence on system size, consistently reaching $\mathcal{D}\approx0.95$--$1$ under VCQA regardless of topology. The ground-subspace probability $\mathcal{P}_{\rm GS}$ behaves differently: it depends more strongly on both $N$ and connectivity, ranging from a narrow, high band for $S_N$ ($0.85$--$0.95$) to a wider, lower one for $L_N$ and $C_N$ ($0.5$--$0.8$, with clear degradation as $N$ grows).

\subsection{Weighted Maxcut}
\begin{figure}[t]
\centering
	\includegraphics[width=1\linewidth]{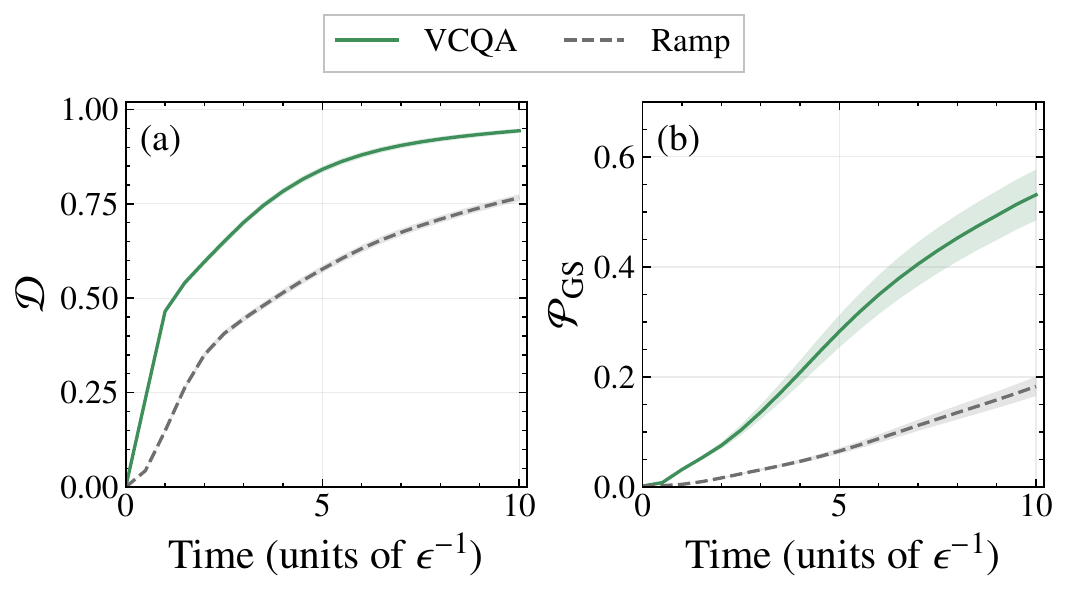}
	\caption{Performance of the VCQA solution as a function of the annealing time for 100 random instances of the weighted maxcut problem (shaded bands indicate the spread across instances). Solid green lines use the VCQA method and dashed grey lines use the standard ramp schedule function. Panel (a) shows $\mathcal{D}$ and panel (b) shows $\mathcal{P}_{\rm GS}$.}
	\label{FigMaxCut}
\end{figure}
While $L_N$, $C_N$, and $S_N$ are physically motivated connectivities, they say little about how VCQA performs on an optimization problem, a hot topic in the quantum computing community. Weighted MaxCut is the standard testbed for such a purpose. We consider 100 random instances of this problem, a fundamental NP-hard problem in graph theory and binary combinatorial optimization, on $N=10$-qubit 5-regular random graphs, i.e., each vertex connected to five others. Consider a graph with edges $E$ and vertices $V$, that is $\mathcal{G}=(V,E)$. Also, we consider a set of weights $w_{j,k}\in [0,1]$, such that each edge corresponds to one weight; in other words, the weight $w_{j,k}$ corresponds to the edge linking vertices $j$ and $k$. The goal of this problem is to divide the vertices $V$ into two disjoint groups such that the total sum of the weights of edges crossing the two groups is maximized. The QUBO form for the cost function of this problem is given by
\begin{equation}
    f(\vec{x})=\sum_{j,k\in E}w_{j,k}\frac{1-x_jx_k}{2}.
    \label{Eq14}
\end{equation}
Figure~\ref{FigMaxCut} compares the VCQA algorithm against the standard ramp schedule function. As shown in Fig.~\ref{FigMaxCut}~(a), $\mathcal{D}$ under VCQA reaches approximately $0.95$ by $t=10\,\epsilon^{-1}$, after an initial rapid rise to about $0.45$ within the first unit of time, while the ramp schedule reaches only about $0.77$ over the same interval. Fig.~\ref{FigMaxCut}~(b) shows a similar pattern for the ground-subspace probability: VCQA reaches $\mathcal{P}_{\rm GS}\approx0.53$ by $t=10\,\epsilon^{-1}$, while the ramp schedule remains below approximately $0.18$. As in the previous benchmarks, we use only two parameters per schedule function, so that the resulting optimization involves only six parameters in total.

\subsection{p-Spin}
The benchmarks considered so far, $L_N$, $C_N$, $S_N$, and weighted MaxCut, are all two-body (QUBO) problems. Many optimization problems of practical interest, however, involve interactions among more than two variables at once (HUBO), which a purely quadratic cost function cannot capture. Finally, we test VCQA on this broader class using a $p$-spin model with $p=4$, whose Hamiltonian reads
\begin{eqnarray}
    H=&&\sum_j \omega_j^{(1)}\sigma_j^z+\sum_{j,k} \omega_{j,k}^{(2)}\sigma_j^z\sigma_k^z+\sum_{j,k,l} \omega_{j,k,l}^{(3)}\sigma_j^z\sigma_k^z\sigma_l^z\nonumber\\
    &&+\sum_{j,k,l,m} \omega_{j,k,l,m}^{(4)}\sigma_j^z\sigma_k^z\sigma_l^z\sigma_m^z.
    \label{Eq15}
\end{eqnarray}
Such Hamiltonians are challenging benchmarks for quantum annealers and can encode nontrivial computational problems such as brute-force factorization~\cite{Guan2024QuantumSciTechnol} and combinatorial
optimization~\cite{Romero2025CommunPhys}.
\begin{figure}[b]
\centering
	\includegraphics[width=1\linewidth]{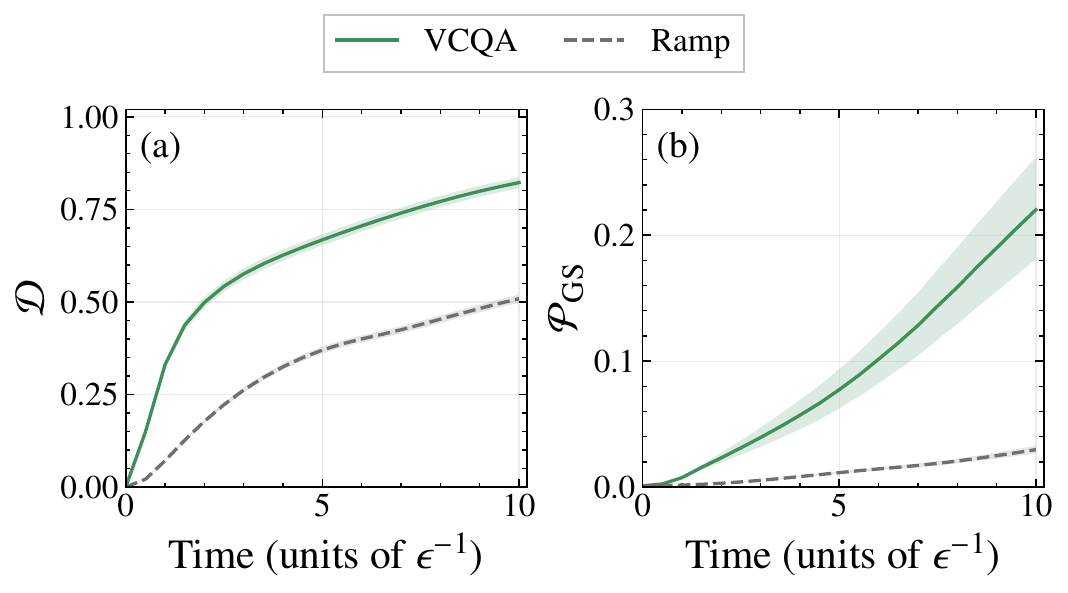}
	\caption{Performance of the VCQA solution as a function of the annealing time for 100 random instances of a ten-qubit, order-4 diagonal $p$-spin model (shaded bands indicate the spread across instances). Solid green lines use the VCQA method and dashed grey lines use the ramp function. Panel (a) shows $\mathcal{D}$ and panel (b) shows $\mathcal{P}_{\rm GS}$.}
	\label{pspin}
\end{figure}

Figure~\ref{pspin} presents results for 100 random ten-qubit instances of the diagonal four-body model, again using only two parameters per schedule function. As shown in Fig.~\ref{pspin}~(a), VCQA reaches $\mathcal{D}\approx0.83$ by $t=10\,\epsilon^{-1}$, compared to approximately $0.51$ for the ramp schedule function. The ground-subspace probability, shown in Fig.~\ref{pspin}~(b), reaches $\mathcal{P}_{\rm GS}\approx0.22$ under VCQA against approximately $0.03$ for the ramp schedule, closely mirroring the weighted-maxcut results of Fig.~\ref{FigMaxCut}. Because the Hamiltonian is diagonal in the computational basis, this benchmark is stoquastic; its distinguishing feature is the presence of multi-spin interactions rather than a sign problem.

\subsection{Non-stoquastic case}
\begin{figure}[t]
\centering
	\includegraphics[width=1\linewidth]{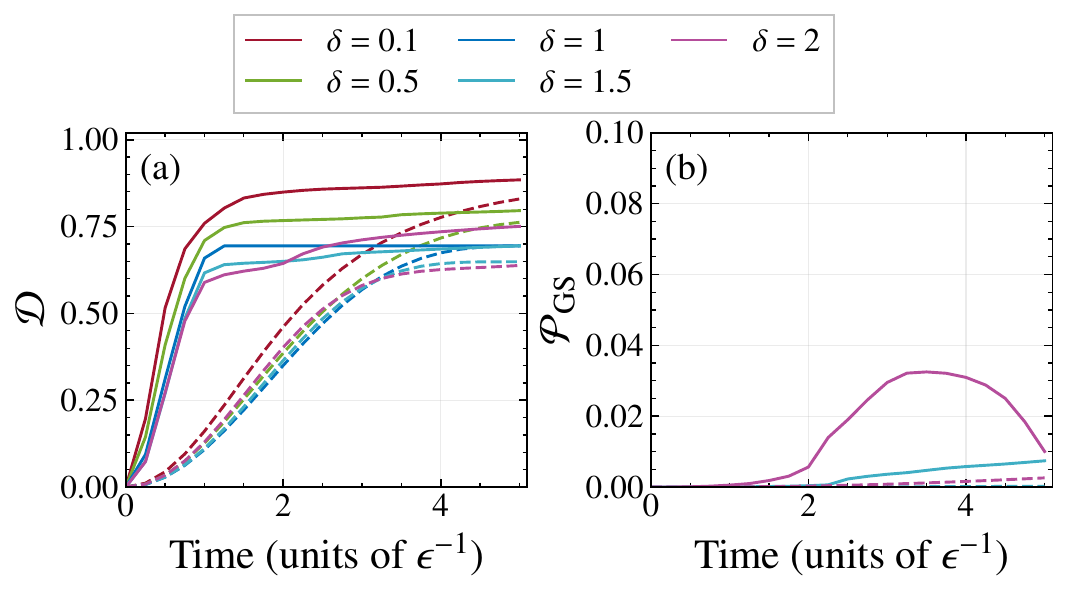}
	\caption{Performance of the VCQA solution as a function of the annealing time for a ten-qubit Heisenberg chain with $y$-axis anisotropy $\delta$ given by Eq.~(\ref{EqH}), for $\delta=0.1,\,0.5,\,1,\,1.5,\,2$. Solid lines use the VCQA method and dashed lines use the ramp function. Panel (a) shows $\mathcal{D}$ and panel (b) shows $\mathcal{P}_{\rm GS}$.}
	\label{Fig06}
\end{figure}
Although available quantum annealers do not allow for the implementation of non-stoquastic Hamiltonians, it is interesting to show the VCQA performance in the presence of non-stoquastic terms. Figure~\ref{Fig06} shows the case of a ten-qubit Heisenberg chain with $y$-axis anisotropy $\delta$. Specifically, the final Hamiltonian corresponds to
\begin{equation}
H_{H}=\omega_H\sum_{j=1}^{N}\sigma^z_j+g_H\sum_{j=1}^{N-1}\left(\sigma_j^x\sigma_{j+1}^x+\delta\sigma_j^y\sigma_{j+1}^y+\sigma_j^z\sigma_{j+1}^z\right) \, ,
\label{EqH}
\end{equation}
with $\omega_H=\epsilon$ and $g_H=0.1\epsilon$, for anisotropy values $\delta=0.1,\,0.5,\,1,\,1.5,\,2$. Figure~\ref{Fig06} compares the VCQA protocol against the ramp schedule function. From Fig.~\ref{Fig06}~(a) we see that VCQA reaches its plateau in $\mathcal{D}$ much faster than the ramp strategy, typically within $t\approx2\,\epsilon^{-1}$, and the two protocols reach comparable final values, between approximately $0.65$ (for $\delta=1,1.5$) and $0.9$ (for $\delta=0.1$), by $t=5\,\epsilon^{-1}$. 

The ground-subspace probability, shown in Fig.~\ref{Fig06}~(b), remains small for both protocols in this non-stoquastic benchmark, in contrast with the Ising cases discussed above: it stays below approximately $0.03$ for all values of $\delta$ and both schedules over the time window considered. VCQA nonetheless reaches consistently larger values of $\mathcal{P}_{\rm GS}$ than the ramp schedule at a given time, and for the largest anisotropy considered, $\delta=2$, $\mathcal{P}_{\rm GS}$ under VCQA rises to a maximum around $t\approx3.5\,\epsilon^{-1}$ before decreasing, a non-monotonic feature not observed for the other values of $\delta$ or for the ramp schedule. This gap between the two metrics is expected for this model: the anisotropic Heisenberg chain is known to exhibit phase transitions, energy-level crossings, and regions of near-degeneracy near the ground state as $\delta$ varies~\cite{AlbarranArriagada2026arXiv}. Near such points, a state can be energetically close to the ground state while remaining nearly orthogonal to it, since $\mathcal{D}$ only tracks the expectation value of $H_f$ and cannot distinguish which state within a near-degenerate manifold was actually reached. The auxiliary-Hamiltonian control therefore accelerates energy minimization even here, but driving the system into the correct state within the ground subspace remains harder than for the stoquastic Ising benchmarks of the previous sections, where no such degeneracy structure is present.

\subsection{Characteristic annealing times}
\label{sec:times}
To quantify how quickly each protocol approaches the ground state, we fit every individual instance $\mathcal{D}(T)$ curve underlying Figs.~\ref{Fig03}--\ref{Fig05}, \ref{FigMaxCut} and \ref{pspin} (the non-stoquastic case of Fig.~\ref{Fig06} is treated separately below) to a decay law in $1-\mathcal{D}(T)$,
\begin{equation}
1-\mathcal{D}(T)=\exp\!\left[-(T/\tau)^{\beta}\right].
\label{Eq17}
\end{equation}

Plotted against $T$, $1-\mathcal{D}(T)$ has the shape of a decaying exponential for both protocols. But the two schedules are built very differently: the ramp is a single, fixed interpolation with nothing to tune, so a single relaxation rate is the natural expectation. VCQA, by contrast, stitches together three independently optimized schedule functions, so its net relaxation plausibly mixes several effective rates rather than one, the classic signature of a stretched exponential. We capture this asymmetry by fitting $1-\mathcal{D}(T)=\exp[-(T/\tau)^{\beta}]$ to every individual instance's curve, with $\beta$ fixed at 1 for the ramp schedule function, so that it reduces to a single-rate exponential, and $\beta$ left free for VCQA. Fits use non-linear least squares on $\mathcal{D}$ directly, with one fit per instance performed before any averaging. In this class of fitting $\beta=1$ corresponds to an ordinary exponential relaxation, $\beta<1$ to a stretched exponential (fast at first, then a dragging tail, the signature of a distribution of local relaxation rates rather than a single one), and $\beta>1$ to a compressed exponential (little change at first, then a comparatively sharp transition).

Table~\ref{TabTimes} reports these fits for $N=10$ qubits across the $L_N$, $C_N$, and $S_N$ connectivities, the weighted MaxCut instances, and the diagonal four-body model. Within each of the three Ising  topologies, $\tau$ is essentially independent of $N$ over the range $N=5$--$10$ studied here (a spread of at most $12\%$ for VCQA and $3\%$ for the ramp across sizes), so the $N=10$ values shown are representative of the whole family.

\begin{table*}[t]
\centering
\begin{tabular}{llcccc}
\hline
Case & Arm & $\tau$ & $\beta$ & $R^2$ & $\tau_{\rm Ramp}/\tau_{\rm VCQA}$ \\
\hline
\multirow{2}{*}{$L_N$}  & VCQA & $1.44\pm0.21$ & $0.67\pm0.09$ & $0.9931$ & \multirow{2}{*}{$2.50\pm0.24$} \\
                        & Ramp & $3.57\pm0.36$ & ---- & $0.9751$ &  \\
\multirow{2}{*}{$C_N$}  & VCQA & $1.45\pm0.23$ & $0.64\pm0.09$ & $0.9932$ & \multirow{2}{*}{$2.43\pm0.24$} \\
                        & Ramp & $3.48\pm0.37$ & ---- & $0.9806$ &  \\
\multirow{2}{*}{$S_N$}  & VCQA & $1.60\pm0.22$ & $0.96\pm0.09$ & $0.9949$ & \multirow{2}{*}{$2.66\pm0.17$} \\
                        & Ramp & $4.25\pm0.46$ & ---- & $0.9956$ &  \\
\multirow{2}{*}{MaxCut} & VCQA & $2.27\pm0.34$ & $0.76\pm0.12$ & $0.9930$ & \multirow{2}{*}{$2.65\pm0.09$} \\
                        & Ramp & $6.02\pm0.90$ & ---- & $0.9736$ &  \\
\multirow{2}{*}{Four-body} & VCQA & $4.47\pm1.46$ & $0.65\pm0.12$ & $0.9835$ & \multirow{2}{*}{$2.99\pm0.68$} \\
                        & Ramp & $12.53\pm2.04$ & ---- & $0.9563$ &  \\
\hline
\end{tabular}
\caption{Decay-law fit of $1-\mathcal{D}(T)=\exp[-(T/\tau)^{\beta}]$ to
individual ten-qubit instances (100 random instances in each case). For ramp we fix $\beta=1$ and $\beta$ left free for VCQA. $\tau$ is in units of
$\epsilon^{-1}$; uncertainties are the standard deviation across
instances. $R^2$ is the goodness of fit of the model actually used. The last
column is the paired, instance-by-instance ratio of characteristic
times $\tau_{\rm Ramp}/\tau_{\rm VCQA}$ (mean $\pm$ sd).}
\label{TabTimes}
\end{table*}

For every case, VCQA's $\beta<1$ reflects a stretched exponential, consistent with a piecewise-optimized schedule that combines several effective local rates along the evolution rather than a single one. The last column of Table~\ref{TabTimes} reports the paired, instance-by-instance ratio of characteristic times $\tau_{\rm Ramp}/\tau_{\rm VCQA}$, which ranges from $2.43$ to $2.99$ across the five cases: VCQA is faster than the ramp schedule function on every one of the paired instances studied here.

\section{Discussion}
In traditional quantum annealing, the Hamiltonian that governs the evolution is given by
\begin{equation}
H(t/T)=[1-g(t/T)]H_i + g(t/T) H_f,
\label{Eq18}
\end{equation}
with $g(0)=0$ and $g(1)=1$, the annealing time is given by~\cite{GarciaPintos2023PhysRevLett}
\begin{equation}
t_f = \frac{\langle H_i \rangle_{t_{f}}+\langle H_f \rangle_{0} - \langle H_f \rangle_{t_{f}}}{i\Tr(\bar{\rho}[H_f,H_i])} .
\label{Eq19}
\end{equation}
Here, $\bar{\rho}$ is the time average of the density matrix during the evolution. This equation relates the final accuracy of the solution, given by the final energy $\langle H_f \rangle_{t_{f}}$, with the time needed to reach it. When we instead consider Hamiltonian evolutions governed by Eq.~(\ref{Eq01}), the relation given by Eq.~(\ref{Eq19}) for the annealing time changes to
\begin{equation}
t_f = \frac{\langle H_i \rangle_{t_{f}}+\langle H_f \rangle_{0} - \langle H_f \rangle_{t_{f}}+\mathcal{R}_{31}(t_f)\langle H_{aux}\rangle_{t_f}+\mathcal{C}}{i\Tr(\bar{\rho}[H_f,H_i])} ,
\label{Eq20}
\end{equation}
where
\begin{equation}
\mathcal{C}=\int_{0}^{t_f}\left[(\langle H_f \rangle - \langle H_i \rangle)\frac{d}{dt}\left(\frac{1}{F_{\Sigma}}\right) - \langle H_{aux} \rangle \frac{d}{dt} \left(\frac{\mathcal{R}_{31}}{F_{\Sigma}}\right)\right]dt .
\label{Eq21}
\end{equation}
Here, $F_{\Sigma}=F_1+F_2$ and $\mathcal{R}_{31}=F_3/F_1$; a detailed derivation is given in Appendix~\ref{app:annealing_time}. We note that $\mathcal{C}=0$ and $\mathcal{R}_{31}(t_f)\langle H_{aux}\rangle_{t_f}=0$ for the case of normal annealing evolution, that is, Hamiltonians of the form Eq.~(\ref{Eq18}), since $F_{\Sigma}=1$ and $F_3=0$ identically.

As the coefficients $\mathcal{C}$ and $\mathcal{R}_{31}(t_f)\langle H_{aux}\rangle_{t_f}$ both depend on the shape of $F_1$, $F_2$, and $F_3$, which are arbitrary, together with the expectation values of the initial, final, and auxiliary Hamiltonian, they open the door to annealing evolutions with a reduced time $t_f$ while maintaining the same accuracy in $\langle H_f \rangle_{t_f}$. For example, in the case $F_{\Sigma}=1$, it is necessary to choose $F_1$ and $F_3$ such that the time derivative of $\mathcal{R}_{31}$ has, as much as possible during the evolution, the same sign as $\langle H_{aux}\rangle$. We also observe that when $\langle H_{aux}\rangle$ is approximately constant in time, the second term of the integral tends to zero. The residual boundary term $\mathcal{R}_{31}(t_f)\langle H_{aux}\rangle_{t_f}$ offers a related, second lever: in our simulations $F_3$ stays modest and tends to vanish before $F_1$ reaches zero, which keeps this term small, but a systematic characterization of its magnitude across schedule families remains open. A more detailed study of both coefficients and their bounds is still needed.

Across the $L_N$, $C_N$, and $S_N$ connectivities, the normalized energy distance $\mathcal{D}$ shows little dispersion and is largely insensitive to increasing $N$, while $\mathcal{P}_{\rm GS}$ behaves quite differently, dropping more sharply as connectivity and system size grow. This asymmetry is consistent with $\mathcal{P}_{\rm GS}$ being a many-body observable: it requires the full state to overlap with the ground subspace, so any energy-level compression that comes with added connectivity degrades it directly, whereas $\mathcal{D}$ only tracks the expectation value of $H_f$ and is comparatively insensitive to how the low-lying spectrum crowds together.

The more modest absolute values of $\mathcal{D}$ and $\mathcal{P}_{\rm GS}$ obtained for the weighted MaxCut instances, relative to the Ising benchmarks, are consistent with our choice of only two variational parameters per schedule function throughout this work. MaxCut couples qubits through inhomogeneous weighted edges, which arguably demands a richer schedule to compensate; increasing $N_j$ for this benchmark is a natural next step to test whether the resulting performance gap narrows.

A qualitative comparison with counter-diabatic (CD) driving and other shortcuts to adiabaticity (STA) clarifies the distinctive role of VCQA. CD methods, either exact or approximate, generally require additional non-stoquastic or multi-body Hamiltonian terms that are difficult to implement with current hardware. Even variational CD approaches often involve complex operator structures and analog controls. By contrast, VCQA restricts the auxiliary terms to local $\sigma_z$ fields, which correspond directly to native $Z$ rotations available on essentially all platforms, and compiles the dynamics into short digital layers. This makes VCQA considerably more practical for near-term realizations. Furthermore, while CD protocols attempt to reproduce the adiabatic trajectory, VCQA directly optimizes a finite-depth digital adiabatic evolution, focusing on performance under realistic circuit constraints.

When compared with other variational and digitized approaches, such as digitized adiabatic quantum computing or QAOA-like ans\"atze, VCQA preserves the annealing-inspired structure while introducing a flexible variational schedule. This provides a broadly applicable ansatz without the need for problem-specific tailoring and keeps the number of variational parameters modest, which in turn reduces optimization overhead. The resulting protocol remains hardware-efficient. Thus, VCQA complements rather than competes with existing approaches, offering a resource-efficient alternative tailored to near-term hardware capabilities.

We emphasize that a systematic quantitative benchmarking against all these methods is beyond the present scope, but our qualitative analysis highlights how VCQA combines broad applicability with practical implementability, providing a complementary tool within the landscape of variational and shortcut-to-adiabaticity protocols.

\section{Conclusion}

We have proposed the VCQA paradigm as a hybrid classical-quantum solution that strictly uses the available coherence time of quantum annealers using current technologies. Our approach has two key aspects: the inclusion of an extra term in the total Hamiltonian that modifies the energy gap, and the use of three parametrized schedule functions. These functions are associated with the initial Hamiltonian, the final Hamiltonian, and the new auxiliary Hamiltonian. In particular, due to experimental considerations, we consider $z$-local terms as our auxiliary Hamiltonian. This choice is advantageous as it is native to almost any physical platform, strongly contributing to the nontrivial enlargement of the energy gap.

We have tested our VCQA algorithm across three different connectivities: linear ($L_N$), cyclic ($C_N$), and star ($S_N$), for systems of up to 10 qubits. In all cases, VCQA drives the normalized energy distance $\mathcal{D}$ to values close to $0.95$--$1$ within a time frame of $10\epsilon^{-1}$, consistently above the ramp schedule function, and reaches substantially higher ground-subspace probabilities $\mathcal{P}_{\rm GS}$, particularly for the $S_N$ connectivity. This timescale corresponds to a few nanoseconds for the current state-of-the-art in superconducting circuit technology~\cite{Johnson2011Nature}. This timescale refers exclusively to the physical duration of a single adiabatic evolution, reaffirming the coherent nature of the evolution; the total runtime of the full protocol also needs to include the classical processing and the optimization routine, which are external to the quantum dynamics. We remark that we have used only two parameters per schedule function in the optimization process, which points to significant potential for further improvement by expanding the number of optimization parameters.

Given these results, we tested VCQA on 100 random instances of the weighted MaxCut problem and on a diagonal $p$-spin model, obtaining better performance than the corresponding ramp schedule in both metrics, although with more modest absolute values of $\mathcal{D}$ and $\mathcal{P}_{\rm GS}$ than in the Ising benchmarks. Fitting a stretched exponential to each individual instance further shows that VCQA's approach to the ground state is genuinely stretched ($\beta<1$) in essentially every case, with a paired, instance-by-instance speed-up over the ramp schedule ranging from a factor of $2.4$ to $3.0$ across all five benchmarks (see Table~\ref{TabTimes}).

We also tested this protocol on a non-stoquastic Heisenberg chain. Even if current experiments may only tackle spin-glass models, it is worth showing the performance of our VCQA proposal on a non-stoquastic case. There, VCQA accelerates the short-time growth of $\mathcal{D}$ relative to the ramp schedule, although the ground-subspace probability $\mathcal{P}_{\rm GS}$ remains small for both protocols within the time window studied, indicating that reaching the true ground subspace of this non-stoquastic model is considerably harder than for the stoquastic Ising benchmarks.

This work points toward a fully coherent use of quantum annealers, using their quantum features to reach faster, more accurate solutions.

\section*{Acknowledgement}

We acknowledge financial support from Agencia Nacional de Investigaci\'on y Desarrollo (ANID): Center for Nanoscience and Nanotechnology, CEDENNA, Project CIA250002, Proyecto de Exploraci\'on grant 13250014, Fondecyt Regular grant 1231172.

\section*{Authorship contribution statement}
N.B.: Conceptualization, Methodology, figures production, Investigation, Writing -- original draft, review, and editing. G.A.B: Conceptualization, Methodology, numerical calculations, Writing -- original draft, review, and editing. E.S.: Conceptualization, Supervision, Writing -- review and editing. F.A.-A.: Conceptualization, Methodology, Supervision, Project administration, Writing -- review and editing.

\section*{Declaration of competing interest}
The authors declare that they have no known competing financial interests or personal relationships that could have appeared to influence the work reported in this paper.

\section*{Data availability}
The numerical data and code required to reproduce the figures will be deposited in a public repository before publication.

\newpage

\bibliographystyle{unsrtnat}
\bibliography{Mybib}

\newpage
\appendix
\onecolumn
\section{Annealing time}
\label{app:annealing_time}

Let us consider the time evolution governed by the Hamiltonian
\begin{equation}
H(t)=F_1(t/t_f) H_i + F_2(t/t_f) H_f + F_3(t/t_f) H_{aux},
\label{A01}
\end{equation}
with $F_1(0)=F_2(1)=1$ and $F_1(1)=F_2(0)=F_3(0)=F_3(1)=0$. Without
loss of generality, we set the zero of energy so that $\langle
H_i\rangle_0=0$, i.e., $H_i\to H_i-\langle H_i\rangle_0\mathbb{1}$;
this shift leaves all commutators in Eqs.~(\ref{A02})--(\ref{A06})
unchanged and simplifies the boundary term below. This convention is
local to the present derivation: elsewhere in this work $H_i=-\epsilon
\sum_j\sigma_j^x$, for which $\langle H_i\rangle_0=-N\epsilon\neq0$.
Then,
\begin{equation}
\frac{d}{dt}\langle H_i \rangle=iF_2\Tr(\rho[H_f,H_i])+iF_3\Tr(\rho[H_{aux},H_i]),
\label{A02}
\end{equation}
\begin{equation}
\frac{d}{dt}\langle H_f \rangle=iF_1\Tr(\rho[H_i,H_f])+iF_3\Tr(\rho[H_{aux},H_f]),
\label{A03}
\end{equation}
\begin{equation}
\frac{d}{dt}\langle H_{aux} \rangle=iF_1\Tr(\rho[H_i,H_{aux}])+iF_2\Tr(\rho[H_f,H_{aux}]).
\label{A04}
\end{equation}

If we consider that $[H_f,H_{aux}]=0$, as in the cases of a $z$-local
auxiliary Hamiltonian for $H_f$ given by a spin-glass one, we can write
\begin{eqnarray}
&&\frac{d}{dt}\langle H_{i} \rangle - \frac{d}{dt}\langle H_{f} \rangle + \mathcal{R}_{31}\frac{d}{dt}\langle H_{aux} \rangle=iF_{\Sigma}\Tr(\rho[H_f,H_i])\nonumber\\
\Rightarrow&& i\Tr(\rho[H_f,H_i])=\frac{1}{F_{\Sigma}}\frac{d}{dt}(\langle H_{i} \rangle - \langle H_{f} \rangle) + \frac{\mathcal{R}_{31}}{F_{\Sigma}}\frac{d}{dt} \langle H_{aux} \rangle \, .
\label{A05}
\end{eqnarray}
Then, through integration, we obtain
\begin{eqnarray}
&&it_f\Tr(\bar{\rho}[H_f,H_i])=\left(\frac{\langle H_{i} \rangle - \langle H_{f}\rangle}{F_{\Sigma}}\right)\Bigg|_{0}^{t_f}  + \frac{\mathcal{R}_{31} \langle H_{aux} \rangle }{F_{\Sigma}}\Bigg|_{0}^{t_f}\nonumber\\ 
&&+ \int_{0}^{t_f}\left[(\langle H_f \rangle - \langle H_i \rangle)\frac{d}{dt}\left(\frac{1}{F_{\Sigma}}\right) - \langle H_{aux} \rangle \frac{d}{dt} \left(\frac{\mathcal{R}_{31}}{F_{\Sigma}}\right)\right]dt,
\label{A06}
\end{eqnarray}
where $\bar{\rho}=\frac{1}{t_f}\int_0^{t_f}\rho \, dt$. Using the
boundary conditions for the functions $F_1$, $F_2$, and $F_3$ together
with the convention $\langle H_i\rangle_0=0$, we obtain
\begin{equation}
t_f = \frac{\langle H_i \rangle_{t_{f}}+\langle H_f \rangle_{0} - \langle H_f \rangle_{t_{f}}+\mathcal{R}_{31}(t_f)\langle H_{aux}\rangle_{t_f}+\mathcal{C}}{i\Tr(\bar{\rho}[H_f,H_i])} \, ,
\label{A07}
\end{equation}
with
\begin{equation}
\mathcal{C}=\int_{0}^{t_f}\left[(\langle H_f \rangle - \langle H_i \rangle)\frac{d}{dt}\left(\frac{1}{F_{\Sigma}}\right) - \langle H_{aux} \rangle \frac{d}{dt} \left(\frac{\mathcal{R}_{31}}{F_{\Sigma}}\right)\right]dt \,  .
\label{A08}
\end{equation}
Note that $\mathcal{R}_{31}(0)=F_3(0)/F_1(0)=0$ identically, so only
the value at $t_f$ survives in Eq.~(\ref{A07}); since $F_1(1)=F_3(1)=0$
simultaneously, $\mathcal{R}_{31}(t_f)$ is formally a $0/0$ limit and is
not assumed to vanish. In practice, $\mathcal{R}_{31}(t_f)\langle
H_{aux}\rangle_{t_f}$ is small, since $F_3$ remains modest throughout
the evolution and tends to vanish faster than $F_1$ approaches zero
near $t_f$; we retain it here rather than assume it vanishes exactly,
as this depends on the specific optimized schedule.

\end{document}